# Unified evaluation of surface-enhanced resonance Raman scattering and fluorescence under strong coupling regime

Tamitake Itoh[1]* and Yuko S. Yamamoto[2]

[1] Nano-Bioanalysis Research Group, Health Research Institute, National Institute of Advanced Industrial Science and Technology (AIST), Takamatsu, Kagawa 761-0395, Japan

[2] School of Materials Science, Japan Advanced Institute of Science and Technology (JAIST), Nomi, Ishikawa 923-1292 Japan

*Corresponding author: tamitake-itou@aist.go.jp

ABSTRACT

Light–matter interaction under strong coupling regime is key in plasmonics. We demonstrate importance of molecular multiple excitons and higher-order plasmons for both enhancement and quenching of resonance Raman and fluorescence of single dye molecule located at plasmonic hotspot under strong coupling regime. The multiple excitons induce complicated spectral changes in plasmon resonance and higher-order plasmons yield drastic quenching for both resonant Raman and fluorescence. A coupled oscillator model composed of plasmon and multiple excitons reproduces the complicated spectral changes. Purcell factors derived from higher-order plasmons reproduce the drastic quenching with considering ultra-fast surface enhanced fluorescence.

Since the discovery of single-molecule (SM) detection of surface-enhanced resonant Raman scattering (SERRS) with enhancement factors around $10^{10}$ to $10^{14}$ by plasmonic nanoparticle (NP) systems [1], electromagnetic (EM) coupling between plasmon and molecular exciton resonance has been investigated to quantitatively clarify the SM SERRS [2-4]. The investigation has elucidated that EM coupling in a field confined by plasmon resonance within several cubic nanometers at junctions, called "hotspots" enables such large enhancement. The investigation under SM SERRS conditions also revealed unique phenomena e.g. strong coupling, in which the EM coupling rates are larger than the dephasing rates of both plasmon and molecular exciton resonances [5-7], ultra-fast surface enhanced fluorescence (ultra-fast SEF), in which the SEF rates overcome the molecular vibrational decay rates, resulting in emission from vibrationally excited states in the electronically excited state [8,9], and others [10].

Therefore, the EM model dealing SERRS is required to correctly evaluate these phenomena under strong coupling conditions. Previous models assumed weak coupling, in which the coupling rate is sufficiently smaller than dephasing rates of both resonances and does not consider ultra-fast SEF [2-4]. Furthermore, we should include in the model not only the quenching of fluorescence, but also the quenching of Raman by higher order plasmons [11]. Such improvement in EM model and experimental examination will resolve several unclear points in experiments e.g. the wide range of maximum enhancement factors from $10^{10}$ to $10^{14}$ [1-4,12].

We developed an EM model to treat the strong coupling between a plasmon and multiple excitons representing the multi-level system of a molecule regarding a Franck-Condon mechanism. The model well reproduces the complicated spectral changes in plasmon resonance of silver NP dimers during SERRS disappearance by only decreasing coupling energy. Purcell’s factor including radiative and nonradiative plasmons successfully reproduces SERRS and SEF with using a relationship between quantum yields of ultra-fast SEF and the coupling energies.

Plasmon resonant light scattering and SERRS spectroscopy of single silver NP dimers were explained in Ref. 5. The average diameter of the silver NPs was 40 nm. NaCl (5 mM) and R6G ($6.4\times10^{-9}$ M) were added to a silver colloidal solution ($1.1\times10^{-10}$ M) for dimerization of NPs including R6G molecules at the dimer junctions or crevasses. The concentration enables near-SM SERRS detections [13]. A green laser light with 2.33 eV and 3.5 W/cm$^2$ was used for SERRS excitation. White light from a 50 W halogen lamp through a dark-field condenser was used to measure the elastic scattering light spectra of the dimers to obtain the plasmon resonance coupled with the molecular resonances.

The model evaluating the strong coupling system composed of a silver NP dimer and a dye molecule locates at the junction of dimer is developed by modifying a coupled-oscillator model, which treats the EM coupling between vacuum EM fluctuation and molecular excitons [14]. A Franck-Condon mechanism indicates that electron-vibration coupling yields multiple excitons [3]. Thus, the coupled-oscillator is composed of an oscillator representing plasmon and multiple oscillators representing molecular excitons. The equations of motion for the coupled oscillators are:

$$\frac{\partial^2 x^p(t)}{\partial t^2}+\gamma^p\frac{\partial x^p(t)}{\partial t}+\omega^{p2}x^p(t)+\sum_{\mathrm{n=1}}^{\mathrm{N}}g_n\frac{\partial x_n^m(t)}{\partial t}=P^p(t) \qquad (1)$$

$$\frac{\partial^2 x_n^m(t)}{\partial t^2}+\gamma_n^m\frac{\partial x_n^m(t)}{\partial t}+\omega_n^{m2}x_n^m(t)-g_n\frac{\partial x^p(t)}{\partial t}=0 \quad (n=1,2,3\ldots) \qquad (2),$$

where $x^p$ and $x_n^m$ are the coordinates of plasmon and $n$-th exciton oscillation, respectively; $\gamma^p$ and $\gamma_n^m$ are the line-widths of plasmon and $n$-th exciton resonance, respectively; $\omega^p$ and $\omega_n^m$ are the resonance frequencies of plasmon and $n$-th exciton, respectively; $g_n$ is the coupling rate between plasmon and $n$-th exciton, and $P^P$ denotes the driving forces representing incident light. We assume that the exciton oscillators are entirely driven by the plasmon oscillator. By assuming that $P^P(t) = P^P e^{-i\omega t}$, where $\omega$ is the incident light frequency, $x^p(t)$ and $x_n^m(t)$ can be derived from Eqs. (1) and (2). In the quasi-static limit, scattering cross-section is $\frac{8\pi}{3}k|\alpha|^2$, where $k = \omega n/c$ is wave vector of light ($n$, $c$, and $\alpha = P^P x^P$ are refractive index of medium, velocity of light, and polarizability, respectively) [14]. By substituting $x_p(t)$ into $\alpha$, one can obtain the cross-section:

$$\frac{8\pi}{3}k|\alpha|^2 \propto \omega^4 \left| \frac{\prod_{n=1}^{N}\left(\omega_n^{m2} - \omega^2 - i\gamma_n^m\omega\right)}{(\omega^2 - \omega^{p2} + i\gamma^p\omega)\prod_{n=1}^{N}(\omega^2 - \omega_n^{m2} + i\gamma_n^m\omega) - \sum_{n=1}^{N}\omega^2 g_n{}^2 \frac{\prod_{k=1}^{N}(\omega^2 - \omega_k^{m2} + i\gamma_k^m\omega)}{(\omega^2 - \omega_n^{m2} + i\gamma_n^m\omega)}} \right|^2$$

$$= C_{sca}(\omega)$$

. (3)

$C_{sca}(\omega)$ represents the scattering spectral shape related to squarer of $Q$ factor of plasmon resonance under $\omega = \omega^P$. Coupling rate $g_n$ is determined by the oscillator strength of electronic transition $f_n$ and the effective mode volume of hotspot $V$:

$$g_n = \left(\frac{1}{4\pi\varepsilon_r\varepsilon_0}\frac{\pi e^2 f_n}{mV}\right)^{1/2}, \quad (4)$$

where $\varepsilon_r$ and $\varepsilon_0$ are relative permittivity 1.77 of a surrounding medium and vacuum one,

respectively, $e$ is elementary charge of electron, and $m$ is free electron mass [15]. $f_n$s are determined from the absorption spectrum of R6G molecules [16]. Figure 1(a) shows the absorption spectrum fitted with quadruple Lorentzian curves representing multiple excitons $f_n$. The inset is the typical SEM image of SERRS active dimer. The fitted spectrum reproduces the experimental one, indicating that the evaluated $f_n$s reasonably represent the multiple excitons of dye molecule. We checked the anti-crossing properties by EM coupling with changing the number of excitons using Eq. (3) with coupling energy $\hbar g_1$ = 200 meV. Figures 1(b1) to 1(b4) illustrate that the anti-crossing behaviors become more complicated with increasing in the excitons. In the single exciton, plasmon resonance spectra clearly exhibits splitting. With increasing in the excitons, the spectra become complicated in the higher energy regions near 2.5 eV without showing the clear spectral splitting. The degree of complexity depending on the number of excitons is caused by multiple strong coupling [17]. We confirm the complexity by comparing experimental changes with calculated ones. Figure 1(c) shows the experimental spectra before and after losing SERRS activity. The blue-shift of plasmon resonance peak from 1.9 to 2.15 eV by losing SERRS activity is observed, but the corresponding red-shift of its counterpart as Fig. 1(b1) is not observed. Figure 1(d) depicts the calculated spectra by decreasing $\hbar g_1$ from 500 to 0 meV. The calculated spectrum with quadruple excitons shows better consistency than that with single excitons, confirming that the complicated spectra without showing a counterpart is caused by strong coupling between plasmon and multiple excitons.

We evaluate the temporal spectral changes in plasmon resonance during losing SERRS activity using the coupled oscillator model. The plasmon resonance spectra commonly exhibit blue-shifts by 100~200 meV simultaneously with the disappearing SERRS activity. The origin of SERRS is EM coupling between plasmon and molecular excitons [2,3]. Thus, this simultaneous blue-shifts are the results of losing coupling energy. Accordingly, we calculated the spectral changes using Eq. (3) by decreasing $\hbar g_n$ with increasing $V$ in Eq. (4). The increase in $V$ means the increase in the effective distance between a molecule and the silver surface by laser induced molecular fluctuation [18]. $\omega^p$ and $\gamma^p$ in Eq. (3) are taken from the plasmon resonance spectra after losing the SERRS activity by assuming that $\hbar g_n = 0$. $\omega_n^m$ and $\gamma_n^m$ are taken from the discussion on Fig. 1(a). The plasmon resonance spectra rapidly blue-shift simultaneously with the SERRS disappearance, indicating the desorption of a molecule from the hotspot. To represent the desorption, we set coupling energy as a sigmoid function against the irradiation time $t$ as $\hbar g_n(t) = \dfrac{\hbar g_n(0)}{1+\exp(t/\tau_a)}$ in Eq. (3), where $\tau_a$ represents the degree of rapidness. Figures 2(a) and 2(b) show the temporal changes in the plasmon resonance spectra and the calculated ones, respectively. The blue-shift and complicated spectral change in the blue region is reproduced in the calculation. Figures 2(c1) and 2(c2) illustrate the experimental spectra before and after losing the SERRS activity and calculated spectra by reducing $\hbar g_1$ from 600 to 0 meV, respectively. The calculated spectra reproduce the blue-shift and the broadening in the experimental ones, indicating that the spectral changes are induced by disappearing multiple-strong coupling with decreasing $\hbar g_n$. We applied this evaluation to twelve dimers and all of them exhibited consistency with the calculations [Figs. SI1-SI12]. The value of $\hbar g_1$ reproducing the experimental spectra are around <600 meV, which corresponds to *V* of a cubic with $(0.76)^3$ $nm^3$ at 2.15 eV, indicating that sub-nanometer cavities are realized at SM SERRS hotspots. Such cavity may be composed of a surface silver atom and a molecule at a hotspot and realize tip-enhanced Raman imaging of molecular internal structures [19,20]. Figure 2(d) shows the anti-crossing behavior calculated with $\hbar g_1$ 400 meV and the plots of the reproduced plasmon resonance peaks by assuming $\hbar g_1$ 400 meV. The positions of all the plots are consistent with the anti-crossing behavior, supporting that the observed spectral changes can be explained by the decreasing in coupling energy during the SERRS disappearance.

We discuss the enhancement and quenching for SERRS and SEF. The reproduction of spectral changes in plasmon resonance by Eq. (3) provides $\hbar g_n$ in Eq. (4). From $\hbar g_n$, one can approximate both enhancement and quenching factors. The mode volume of hotspot represented by $V$ in Eq. (4) is derived from the coupling energies i.e. $\hbar g_1$. From $V$, an EM enhancement factor is evaluated as $F_R = \eta_p F$, where $\eta_p$ and $F$ are the quantum efficiency of plasmon resonant light scattering and Purcell's factor of a dipole plasmon, respectively. $\eta_p$ is estimated to be 5% by the $Q$ factors of experimental plasmon resonance around 10 [21]. Purcell's factor is expressed as $F = \frac{3Q(\lambda_{sc}/n)^3}{4\pi^2 V}$, where $\lambda_{sc}$ and $n$ are emission wavelength and $\varepsilon_r^{1/2}$, respectively [22]. Then, we evaluate the quenching factors $F_{NR}$ using Ref. 11. When a molecule emits light to a free space through a radiative plasmon mode, the light is also quenched by a metal surface through nonradiative plasmon modes. $F_{NR}$ is expressed as a product of an original molecular radiative efficiency and a series of Purcell's factors of nonradiative plasmon modes. We phenomenologically approximate $F_{NR}$ for a molecule close to a NP surface by effective distance $d_{eff}$ in term of large difference between the molecular size ~0.5 nm and NP diameter ~40 nm [23]. Note that in the exact case we should calculate $F_{NR}$ using geometry of hotspots [3]. The $d_{eff}$s are estimated as $V^{1/3}$s. Using $d_{eff}$, Purcell's factors of the plasmon modes are expressed as a series of $F_l$ ($l$ = 1,2,3…):

$$F_l = \frac{3\pi\varepsilon_r (l+1)^2 \omega_{\rm sc} \dfrac{\gamma_l / 2\pi}{(\omega-\omega_l)^2 + \gamma_l{}^2/4}}{4} \left(\frac{\lambda_{\rm sc}}{2\pi a}\right)^3 \left(\frac{a}{a+d_{\rm eff}}\right)^{2l+4}, (5)$$

where $a$ is radius of NP ~20 nm and $\omega_l$ is derived from $l\varepsilon_M(\omega_l)+(l+1)\varepsilon_r=0$ [11]. $\varepsilon_M$ is dielectric function of silver [24] and $\gamma_l$ is plasmon damping rate fixed as an inverse of 10 fs [11]. $F_{\rm NR}$ is presented as follows:

$$F_{\rm NR}\left(\omega_{\rm sc},\eta_{\rm R}\right) \approx \eta_{\rm R} \sum_{l=1}^{\infty} F_l(\omega_{\rm sc}), (6)$$

where $\eta_{\rm R}$ is the quantum efficiency of fluorescence or Raman. "$\approx$"in Eq. (6) indicates one-wavelength approximation [8,11,23]. $\eta_{\rm R}$ of fluorescence is $\eta_{\rm R} = \frac{\Gamma_{\rm R}}{\Gamma_{\rm R}+\Gamma_{\rm NR}}$, where $\Gamma_{\rm R}$ and $\Gamma_{\rm NR}$ are the radiative and nonradiative decay rates, respectively. $\eta_{\rm R}$ = 0.99 is suitable for fluorescent dyes, such as R6G. Raman is a scattering process; thus, $\eta_{\rm R}$ is determined by the competition between dephasing of an excited dipole and Raman radiation like $\eta_{\rm R} = \frac{\Gamma_{\rm RR}}{\Gamma_{\rm RR}+T_2}$, where $\Gamma_{\rm RR}$ and $T_2$ are Raman radiation and dephasing rates, respectively [11]. Thus, $\eta_{\rm R}$ ~0.001 is reasonable for the resonance Raman by considering the ratio of the linewidth of absorption spectrum ~$\text{fs}^{-1}$ and that of Raman spectrum ~$\text{ps}^{-1}$ under a resonant condition.

Figure 3(a) shows the relationship between $1/F_{\rm NR}$ and $F_{\rm R}$ by changing $\hbar g_1$. For $F_{\rm R}$, we approximate $\eta_{\rm p}$ = 0.05 and do not consider the effect of $Q$ factor as $F = \frac{3(\lambda_{\rm sc}/n)^3}{4\pi^2 V}$. $1/F_{\rm NR}$ decreases with increasing in $\eta_{\rm R}$, indicating that the fluorescence obtains drastic quenching compared with Raman. Figure 3(a) also shows that the resonant Raman obtain more quenching than non-resonant Raman, because $\eta_R$ for resonant Raman is larger than that for non-resonant one. This fact may be the reason for that the reported maximum enhancement factor of SERRS ~$10^{10}$ is smaller than that for non-resonant SERS ~$10^{14}$ [1-4,12]. Figure 3(b) shows total enhancement factors $F_{total}$s calculated by $F_{total}(\omega_{ex},\omega_{em},\eta_R)=F_R(\omega_{ex})F_R(\omega_{em})/F_{NR}(\omega_{em},\eta_R)$ excluding the $Q$ factors, where $\hbar\omega_{ex}$ and $\hbar\omega_{em}$ are the excitation and emission photon energies fixed as 2.33 and 2.15 eV, respectively. $F_{total}$ is insensitive to the coupling energy in the region >100 meV, because the increase in $F_R \times F_R$ is canceled by the decrease in $F_{NR}$. Thus, the SERRS and SEF spectra are expected to exclusively depend on the plasmon resonance spectral shapes reflecting the $Q$ factors. In the region $\hbar g_1$ <50 meV, $F_{total}$ for $\eta_R$ < 0.001 decreases, but $F_{total}$ for $\eta_R$ >0.01 increases. This opposite tendency explains the appearance of SEF stronger than SERRS just before the SERRS disappearance [18]. To obtain a more quantitative picture, SERRS cross-section $\sigma_{SERRS}$ (or SEF one $\sigma_{SEF}$) are estimated as a product of $F_{total}$ and resonance Raman cross-section $\sigma_R$ (or fluorescence one $\sigma_F$) by assuming $Q$ = 10. Figures SI13(a) and SI13(b) exhibit $\sigma_R$ and $\sigma_F$, respectively. As regards the two-fold enhancement as $F_{total}Q^2$ [2-4], one can obtain $\sigma_{SERRS}$ ~$10^{-15}$ cm$^2$/meV and $\sigma_{SEF}$ ~$10^{-16}$ cm$^2$/meV in the region of $\hbar g_1$ <100 meV. These values are roughly consistent with the reported values [3,19].

We finally examine the spectral changes in SERRS and SEF during their disappearance process using the coupling oscillator model and both the enhancement and quenching factors. Figure 4(a) shows the temporal changes in experimental SERRS and SEF spectra of a dimer identical to that in Fig. 2(a). Figures 2(a) and 4(a) indicate that the envelope of SEF spectra shows slow blue-shift as well as plasmon resonance spectra, then finally disappears with rapid blue-shift. We reproduce the spectral changes as a function of the coupling energy. Figure 4(b) show the SERRS and SEF spectra extracted from Fig. 4(a). Such blue-shifts are clearly observed. The effect of plasmon resonance spectral shape on enhancement of SERRS and SEF is the $Q$ factor and the spectral modulation [25]. These two are included in the calculation by taking a square root of $C_{\mathrm{sca}}$ in Eq. (3) as $L=\sqrt{C_{sca}}$, because $C_{\mathrm{sca}}$ is proportional to $Q^2$ [25]. Figure SI14(a) shows the combined spectra of SERRS and SEF calculated as a sum of $L(\omega_{\mathrm{ex}})L(\omega_{\mathrm{em}})F_{\mathrm{total}}(\omega_{\mathrm{ex}},\omega_{\mathrm{em}},0.001)\sigma_{\mathrm{R}}(\omega_{\mathrm{ex}},\omega_{\mathrm{em}})$ and $L(\omega_{\mathrm{ex}})L(\omega_{\mathrm{em}})F_{\mathrm{total}}(\omega_{\mathrm{ex}},\omega_{\mathrm{sc}},0.99)\sigma_{\mathrm{F}}(\omega_{\mathrm{ex}},\omega_{\mathrm{em}})$. By comparing Figs. 4(b) and SI14(a), the calculated spectra underestimate the SEF intensity. The underestimation may be due to the use of $\eta_{\mathrm{R}}$ ~ 0.99 for SEF, because ultra-fast SEF occurs for the region $\hbar g_1$ >20meV corresponding to $F_{\mathrm{R}}(\omega_{\mathrm{sc}})>10^4$. $F_{\mathrm{R}}(\omega_{\mathrm{sc}})>10^4$ makes the radiative decay rate of SEF larger than the vibrational relaxation rate ~$10^{12}$ $\mathrm{s}^{-1}$ [8,9]. Figures SI15(a) and SI15(b) explain difference between conventional SEF and ultra-fast SEF [9]. Ultra-fast SEF may have an intrinsically small $\eta_R$ than that of conventional one, because such spectral component is negligible in conventional fluorescence. Thus, as a test we set $\eta_R$ = 0.1 and calculate SEF spectra. By comparing Figs. 4(b) and SI14(b), the calculated SEF spectra is consistent to the experimental ones around $\hbar g_1$ = 500 meV. However, SEF intensity is overestimated around $\hbar g_1$ = 5 meV corresponding to $F_R(\omega_{sc}) < 10^4$. The overestimation is due to the decrease in $F_{NR}$ and indicates that $\eta_R$ = 0.1 is no more suitable for reproducing the SEF spectra in such small coupling energy region because of returning from ultra-fast SEF ($\eta_R$ = 0.1) to conventional SEF ($\eta_R$ = 0.99). Thus, we increase $\eta_R$ with decreasing in $\hbar g_n$ to reproduce experimental spectra. Figure 4(c) shows the calculated SERRS and SEF spectra with increasing $\eta_R$ from 0.1 to 0.99 with decreasing in $\hbar g^1$ from 600 to 5 meV. These SERRS and SEF spectra are correctly reproduced in the calculations. We applied this method to twelve dimers and all the experimental spectra are well reproduced [Fig. SI16-27]. The experimental cross-sections are consistent with the calculations within a factor of ~10. This value is acceptable considering the huge enhancement factor of $10^{10}$ involved in SERRS. Figure 4(d) shows the relationship between $\hbar g_1$ and $\eta_R$ of SEF for the reproductions. It is clearly observed that the increase in $\eta_R$ with the decrease in $\hbar g_1$, indicating the returning from ultra-fast SEF to conventional SEF.

In this work, we analyzed SERRS and SEF under strong coupling conditions between plasmon and molecular excitons. The analysis method is composed of the following: 1) evaluation of coupling energies between plasmon and multiple excitons using a coupled oscillator model, 2) derivation of mode volumes from the energies, 3) estimation of enhancement and quenching factors using Purcell factors by the volumes, and 4) reproduction of SERRS and SEF spectra using these factors. The coupled oscillator is composed of an oscillator representing plasmon and quadruple oscillators representing multiple excitons. The model revealed the complicated spectral changes as multiple-strong coupling. The derived enhancement and quenching factors reproduced experimental SERRS and SEF spectra by considering returning from ultra-fast SEF to conventional SEF. We plan to apply this method to various plasmon-enhanced spectroscopies with unique hotspots [25,26].

**Figure captions**

Fig. 1 (a) Absorption spectrum of aqueous solution of R6G (red curves) fitted with four Lorentzian curves (dashed curves). $\omega_n^m$ and $\gamma_n^m$ are indicated in the panel. Black curve is a sum of Lorentzian curves. $f_n$ are indicated in the panel [16]. Insets: typical SEM image of silver NP dimer showing SERRS activity. Scale bar is 50 nm. (b)

Anti-crossing properties appearing in plasmon resonance spectra calculated by Eq. (3) under the conditions with e.g. $\hbar g_1$ = 200 meV with (b1) single exciton; (b2) double excitons; (b3) triple excitons; and (b4) quadruple excitons. (c) Experimental plasmon resonance spectra before (red curve) and after (black curve) losing SERRS activity. (d) Calculated spectra by Eq. (3) with i.e. $\hbar g_1$ = 500 meV using single exciton (dashed red curve) and quadruple excitons (red curve) and not using exciton coupled with the plasmon (black curve).

FIG. 2 (a) Images of temporal changes in plasmon resonance spectra during a SERRS disappearance. White bar indicates loss by laser notch filter. (b) Images of changes in plasmon resonance spectra by Eq. (3) by decreasing $\hbar g_1$ from 600 to 0 meV with $\tau_a$ = 13.3. (c1) Experimental plasmon resonance spectra of dimer before (red curve) and after (black curve) losing SERRS activity. (c2) Plasmon resonance spectra calculated by Eq. (3) with $\hbar g_1$ = 600 meV (red curve) and with $\hbar g_1$ = 0 meV (black curve). Anti-crossing properties of strong coupling between plasmon and quadruple excitons calculated by Eq. (3) for $\hbar g_1$ = 400 meV with plasmon resonance peak positions reproduced by Eq. (3) for twelve dimers (open circles) in Figs. SI1-SI12.

FIG. 3 (a) $\hbar g_1$ dependence of $1/F_{\mathrm{NR}}$ (black curves with open circles) calculated by Eq. (6) with $\eta_{\mathrm{R}}$ from 0.00001 (upper) to 0.99 (bottom) per a factor of 10 and $\hbar g_1$ dependence of $F_{\mathrm{R}}$ at 2.33 eV (blue curve with open circles) and 2.15 eV (red curve with open circles). (b) $\hbar g_1$ dependence of $F_{\mathrm{total}}(2.33, 2.15, \eta_{\mathrm{R}}) = F_{\mathrm{R}}(2.33) F_{\mathrm{R}}(2.15) / F_{\mathrm{NR}}(2.15, \eta_{\mathrm{R}})$ with $\eta_{\mathrm{R}}$ from 0.00001 (upper) to 0.99 (bottom) per a factor of 10. Upper axis indicates $d_{\mathrm{eff}}$ derived from $V^{1/3}$.

FIG. 4 (a) Image of temporal changes in SERRS and SEF spectra for the NP dimer identical to Fig. 2(a). White bar indicates loss by laser notch filter. (b) SERRS and SEF spectra extracted from (a) at detection times indicated in the panels. Maxima of envelopes are indicated by arrows. (c) Combined spectra of SERRS and SEF calculated with a product of $L(2.33)L(\omega_{\mathrm{em}})F_{\mathrm{total}}(2.33, \omega_{\mathrm{em}}, \eta_{\mathrm{R}})$ and $\sigma_{\mathrm{R}}$ plus $\sigma_{\mathrm{F}}$. The values of $\hbar g_1$ and $\eta_{\mathrm{R}}$ for SEF are indicated in the panels. $\eta_{\mathrm{R}}$ for SERRS is fixed 0.001. Maxima of envelopes are indicated by the arrows. (d) $\hbar g_1$ dependence of $\eta_{\mathrm{R}}$ of SEF used for Figs. SI16-SI27.

Fig.1

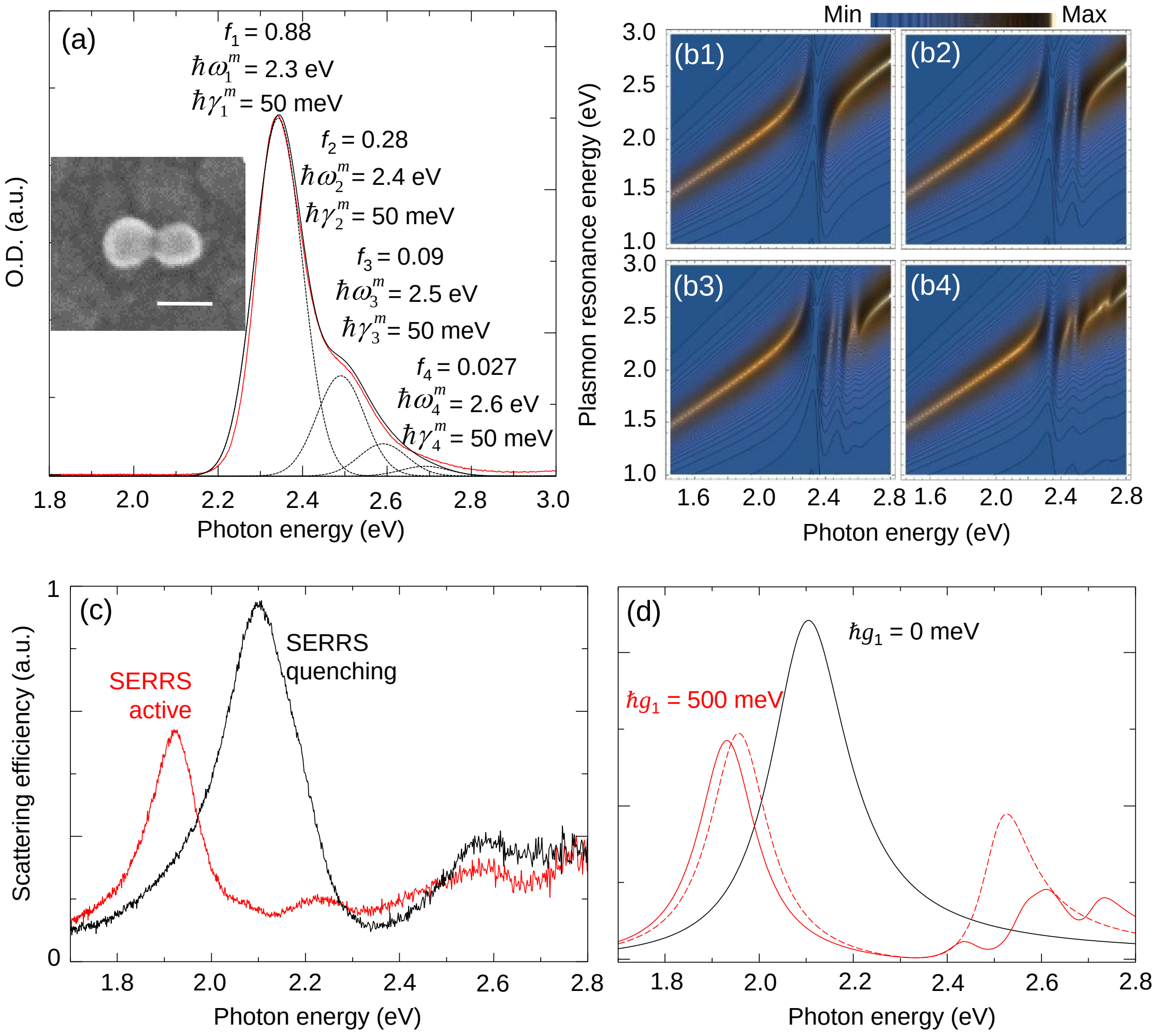

(a)
$f_1$ = 0.88
$\hbar\omega_1^m$ = 2.3 eV
$\hbar\gamma_1^m$ = 50 meV
$f_2$ = 0.28
$\hbar\omega_2^m$ = 2.4 eV
$\hbar\gamma_2^m$ = 50 meV
$f_3$ = 0.09
$\hbar\omega_3^m$ = 2.5 eV
$\hbar\gamma_3^m$ = 50 meV
$f_4$ = 0.027
$\hbar\omega_4^m$ = 2.6 eV
$\hbar\gamma_4^m$ = 50 meV
O.D. (a.u.)
1.8 2.0 2.2 2.4 2.6 2.8 3.0
Photon energy (eV)
Min Max
(b1) (b2) (b3) (b4)
Plasmon resonance energy (eV)
3.0 2.5 2.0 1.5 1.0
1.6 2.0 2.4 2.8
Photon energy (eV)
(c)
SERRS active
SERRS quenching
Scattering efficiency (a.u.)
1 0
1.8 2.0 2.2 2.4 2.6 2.8
Photon energy (eV)
(d)
$\hbar g_1$ = 0 meV
$\hbar g_1$ = 500 meV
1.8 2.0 2.2 2.4 2.6 2.8
Photon energy (eV)

Fig. 2

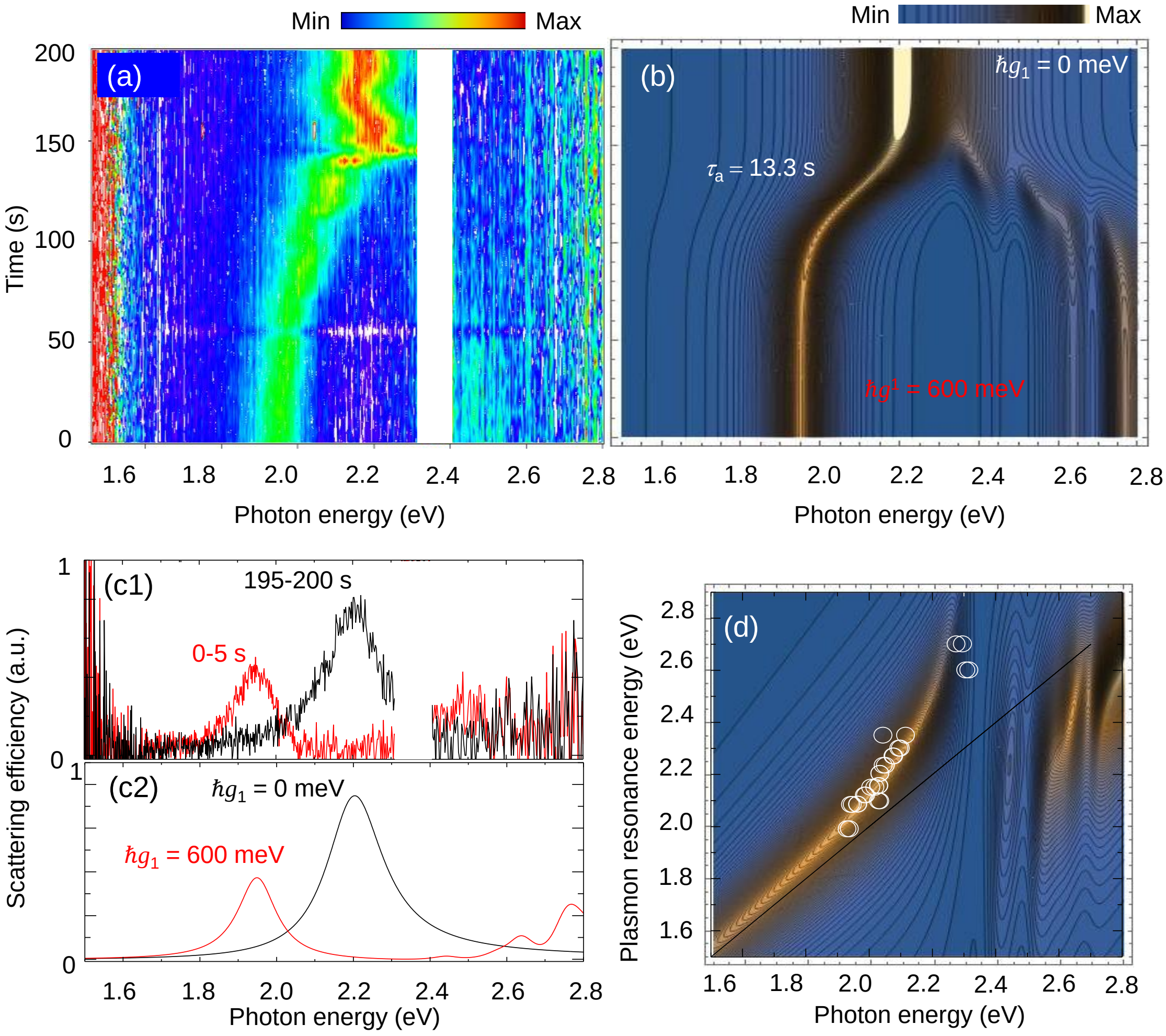

Fig. 3

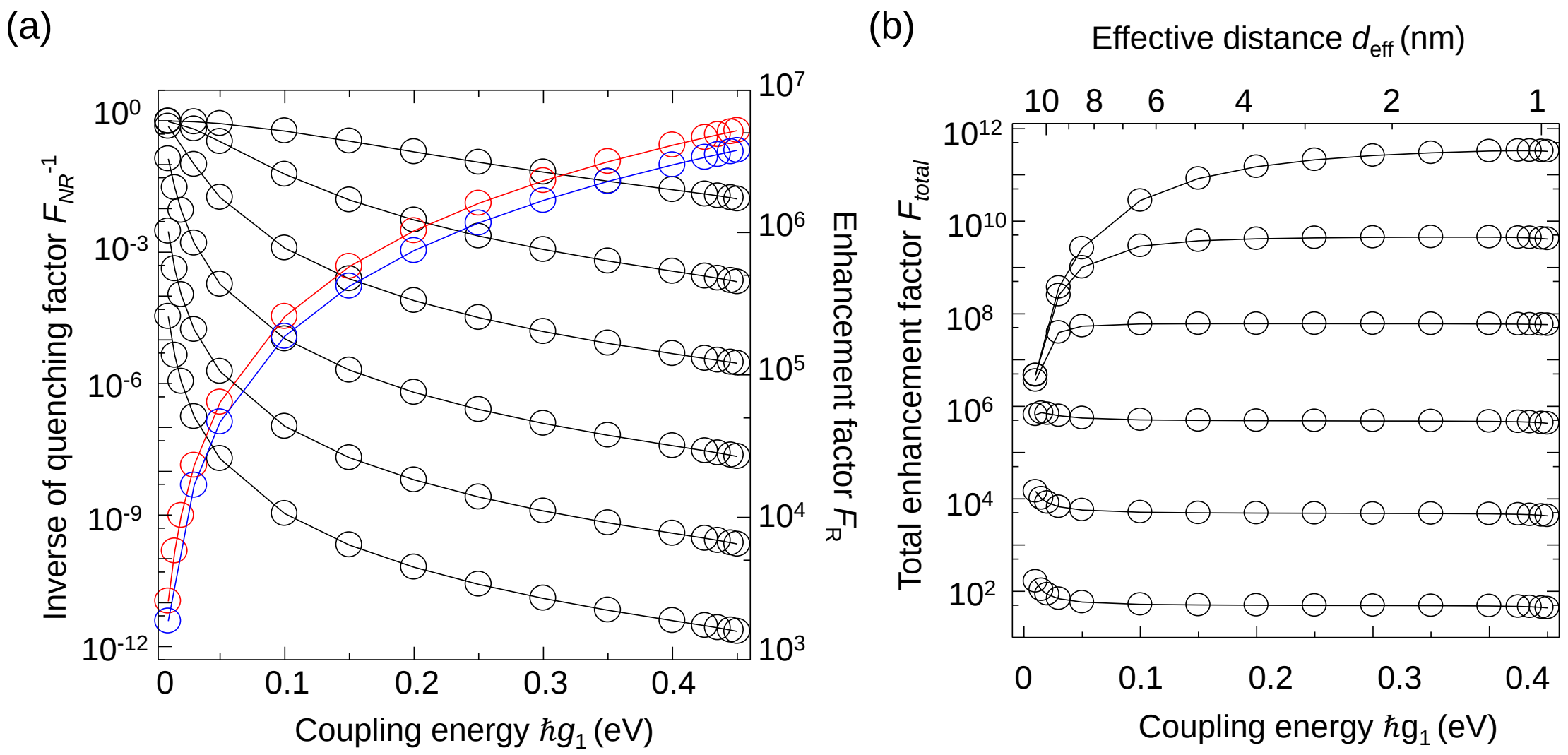

Fig. 4

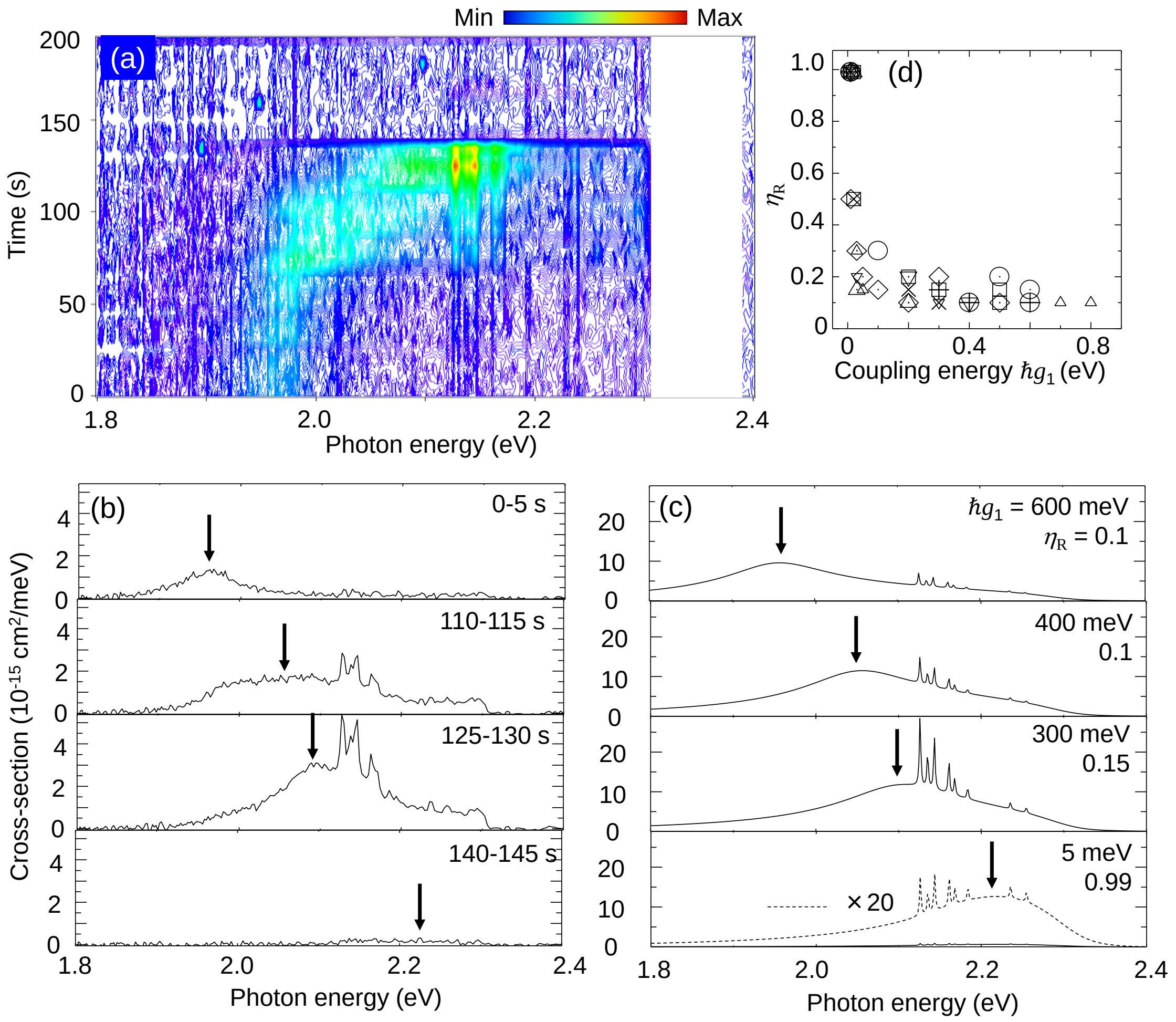